\documentclass[12pt]{iopart}
\usepackage{amssymb}
\usepackage{amsthm}
\usepackage{graphicx}
\usepackage{graphics,epsfig,color}

\newcommand{\bfx}{{\bf x}}
\newcommand{\bfxp}{{{\bf x}^\prime}}
\newcommand{\wbfx}{{\widehat{\bf x}}}
\newcommand{\wbfxp}{{\widehat{\bf x}}^\prime}
\newcommand{\N}{{\mathbf N}}

\newcommand{\R}{{\mathbf R}}
\newcommand{\Si}{{\mathbf S}}
\newcommand{\Hi}{{\mathbf H}}
\newcommand{\h}{{\mathfrak h}}
\newcommand{\g}{{\mathfrak g}}

\newcommand{\Z}{{\mathbf Z}}
\newcommand{\C}{{\mathbf C}}

\newcommand{\mcI}{{\mathcal I}}
\newcommand{\mch}{{\mathcal H}}
\newcommand{\mcg}{{\mathcal G}}

\newtheorem{lemma}{Lemma}[section]
\newtheorem{thm}[lemma]{Theorem}

\begin{document}

\title[
Fourier and Gegenbauer expansions for a fundamental solution of $-\Delta$ in $\Hi^d$]{Fourier and Gegenbauer expansions for a fundamental
solution of the Laplacian in the hyperboloid model of hyperbolic geometry }

\author{H S Cohl${}^{1,2}$ and E G Kalnins${}^3$}

\address{$^1$Information Technology Laboratory, National Institute of Standards and Technology, Gaithersburg, MD, USA}
\address{$^2$Department of Mathematics, University of Auckland, Auckland, New Zealand}
\address{$^3$Department of Mathematics, University of Waikato, Hamilton, New Zealand}
\ead{hcohl@nist.gov}
\begin{abstract}
Due to the isotropy $d$-dimensional hyperbolic space, there exist a spherically 
symmetric fundamental solution for its corresponding Laplace-Beltrami operator.  
On the $R$-radius hyperboloid model of $d$-dimensional hyperbolic geometry with $R>0$
and $d\ge 2$, we compute azimuthal Fourier expansions for a fundamental solution of Laplace's 
equation.  For $d\ge 2$, we compute a Gegenbauer polynomial expansion in geodesic polar 
coordinates for a fundamental solution of Laplace's equation on this negative-constant 
sectional curvature Riemannian manifold.  In three-dimensions, an addition theorem for the azimuthal 
Fourier coefficients of a fundamental solution for Laplace's equation is obtained through 
comparison with its corresponding Gegenbauer expansion.
\end{abstract}

% Keywords: Hyperbolic geometry; Fundamental solution; Laplace's equation; Fourier series; Gegenbauer series; Separation of variables
\maketitle

\section{Introduction}
\label{Introduction}

In this paper we discuss eigenfunction expansions for a fundamental
solution of Laplace's equation in the hyperboloid model of $d$-dimensional 
hyperbolic geometry.
In particular, for a fixed $R\in(0,\infty)$ and $d\ge 2$, 
we derive and discuss 
Fourier cosine and Gegenbauer polynomial expansions in rotationally invariant 
coordinate systems, for a previously derived (see Cohl \& Kalnins (2011) \cite{CohlKalI})
spherically symmetric fundamental solution of the 
Laplace-Beltrami operator in the hyperboloid model of hyperbolic geometry.
Useful background material relevant for this paper can be found in 
Vilenkin (1968) \cite{Vilen}, 
Thurston (1997) \cite{Thurston}, Lee (1997) \cite{Lee} and 
Pogosyan \& Winternitz (2002) \cite{PogWin}.

This paper is organized as follows.  In section \ref{Globalanalysisonthehyperboloid}, for the 
hyperboloid model of $d$-dimensional hyperbolic geometry, we describe some of its global properties, 
such as its geodesic distance function, geodesic polar coordinates, 
Laplace-Beltrami operator (Laplacian), radial harmonics, and several previously 
derived equivalent expressions for a radial fundamental solution of Laplace's equation.  In 
section \ref{FourierexpansionsforaGreensfunctioninthehyperboloidmodel}, for $d\ge 2$,
we derive and discuss Fourier cosine series for a fundamental solution of Laplace's equation 
about an appropriate azimuthal angle in rotationally invariant coordinate systems, and show 
how the resulting Fourier coefficients compare to the those in Euclidean space.
In section \ref{Gegenbauerexpansioninhyperbolichypersphericalcoordinates}, for $d\ge 2$, 
we compute Gegenbauer polynomial expansions in geodesic polar coordinates,
for a fundamental solution of Laplace's equation in the hyperboloid model of hyperbolic geometry.
In section \ref{Discussion} we discuss possible directions of research in this area.

Throughout this paper we rely on the following definitions.  
For $a_1,a_2,\ldots\in\C$, if $i,j\in\Z$ and $j<i$ then
$\sum_{n=i}^{j}a_n=0$ and $\prod_{n=i}^ja_n=1$.
The set of natural numbers is given by $\N:=\{1,2,\ldots\}$, the set
$\N_0:=\{0,1,2,\ldots\}=\N\cup\{0\}$, and the set
$\Z:=\{0,\pm 1,\pm 2,\ldots\}.$  The set $\R$ represents the real numbers.
%, the set of integers $\Z:=\{0,\pm 1,\pm 2,\ldots\},$
%the set of real numbers $\R:=(-\infty,\infty)$, the set of complex numbers 
%$\C:=\{a+bi:a,b\in\R\}$.

\section{Global analysis on the hyperboloid}
\label{Globalanalysisonthehyperboloid}

\subsection{The hyperboloid model of hyperbolic geometry}

Hyperbolic space, developed independently by Lobachevsky and Bolyai around 1830 
(see Trudeau (1987) \cite{Trudeau}), is a fundamental example of a space 
exhibiting hyperbolic geometry.  Hyperbolic geometry is analogous to 
Euclidean geometry, but such that Euclid's parallel postulate is no longer assumed to hold.
There are several models of $d$-dimensional hyperbolic space $\Hi_R^d$, including the 
Klein, Poincar\'{e}, hyperboloid, upper-half space and hemisphere models
(see Thurston (1997) \cite{Thurston}).  The hyperboloid model for $d$-dimensional 
hyperbolic geometry (hereafter referred to as the hyperboloid model, or more simply, 
the hyperboloid), is closely related to the Klein and Poincar\'{e}
models: each can be obtained projectively from the others.  
The upper-half space and hemisphere models can be obtained from 
one another by inversions with the Poincar\'{e} model 
(see section 2.2 in Thurston (1997) \cite{Thurston}).
The model of hyperbolic geometry which we will be focusing on in this paper, is the hyperboloid model.

Minkowski space $\R^{d,1}$ is a $(d+1)$-dimensional pseudo-Riemannian manifold
which is a real finite-dimensional vector space, with Cartesian coordinates given by $\bfx=(x_0,x_1,\ldots,x_d)$. 
The hyperboloid model, also known as the Minkowski or Lorentz models, represents points in 
this space by the upper sheet $(x_0>0)$ of a two-sheeted hyperboloid embedded in 
the Minkowski space $\R^{d,1}$.  
It is equipped with a nondegenerate, symmetric bilinear form, the Minkowski bilinear form
\[
[\bfx,{\mathbf y}]=x_0y_0-x_1y_1-\ldots-x_dy_d.
\]
The above bilinear form is symmetric, but not positive-definite, so it is 
not an inner product.  It is defined analogously with the Euclidean inner product 
for $\R^{d+1}$
\[
(\bfx,{\mathbf y})=x_0y_0+x_1y_1+\ldots+x_dy_d.
\]
The variety $[\bfx,\bfx]=x_0^2-x_1^2-\ldots-x_d^2=R^2$, for $\bfx\in\R^{d,1}$, using the language of 
Beltrami (1869) \cite{Beltrami} (see also p.~504 in Vilenkin (1968) \cite{Vilen}),
defines a pseudo-sphere of radius $R$. Points on the 
pseudo-sphere with zero radius coincide with a cone.  
Points on the pseudo-sphere with radius greater than zero 
lie within this cone, and points on the pseudo-sphere
with purely imaginary radius lie outside the cone.

For a fixed $R\in(0,\infty),$ the $R$-radius hyperboloid model is a 
maximally symmetric, simply connected, $d$-dimensional Riemannian manifold 
with negative-constant sectional curvature (given by $-1/R^2$, see 
for instance p.~148 in Lee (1997) \cite{Lee}), whereas Euclidean 
space $\R^d$ equipped with the Pythagorean norm, is a Riemannian manifold with zero 
sectional curvature.  The hypersphere $\Si^d$, is an example of a space (submanifold) with 
positive-constant sectional curvature (given by $1/R^2$).

In our discussion of a fundamental solution for Laplace's equation
in the hyperboloid model $\Hi_R^d$, we focus on the positive radius 
pseudo-sphere which can be parametrized through 
{\it subgroup-type coordinates,} i.e. those which correspond to a maximal
subgroup chain $O(d,1)\supset \ldots$ (see for instance Pogosyan \& Winternitz (2002) \cite{PogWin}).
There exist separable coordinate systems which parametrize points on the
positive radius pseudo-sphere (i.e. such as those which are analogous to
parabolic coordinates, etc.) which can not be constructed using maximal 
subgroup chains (we will no longer discuss these).

Geodesic polar coordinates are coordinates which correspond 
to the maximal subgroup chain given by $O(d,1)\supset O(d)\supset \ldots$.
What we will refer to as {\it standard geodesic polar coordinates} 
correspond to the subgroup chain given by 
$O(d,1)\supset O(d)\supset O(d-1)\supset \cdots \supset O(2).$
Standard geodesic polar coordinates
(see Olevski{\u\i} (1950) \cite{Olevskii};
Grosche, Pogosyan \& Sissakian (1997) \cite{groschepogsis}), similar 
to standard hyperspherical coordinates in Euclidean space, can be given by
\begin{equation}
\left.
\begin{array}{rcl}
x_0&=&R\cosh r\\[0.02cm]
x_1&=&R\sinh r\cos\theta_1\\[0.1cm]
x_2&=&R\sinh r\sin\theta_1\cos\theta_2\\[0.1cm]
&\vdots&\\[0.02cm]
x_{d-2}&=&R\sinh r\sin\theta_1\cdots\cos\theta_{d-2}\\[0.1cm]
x_{d-1}&=&R\sinh r\sin\theta_1\cdots\sin\theta_{d-2}\cos\phi\\[0.1cm]
x_{d}&=&R\sinh r\sin\theta_1\cdots\sin\theta_{d-2}\sin\phi,
\end{array}
\quad\right\}
\label{standardhyp}
\end{equation}

\noindent where $r\in[0,\infty)$, $\phi\in[0,2\pi)$, 
and $\theta_i\in[0,\pi]$ for $i\in\{1,\ldots,d-2\}$.  

In order to study fundamental solutions on the hyperboloid, we need to describe how 
one computes distances in this space.  One may naturally compare distances on the 
positive radius pseudo-sphere through analogy with the $R$-radius hypersphere.  
Distances on the hypersphere are simply given by arc lengths, angles between two 
arbitrary vectors, from the origin, in the ambient Euclidean space.
We consider the $d$-dimensional hypersphere embedded in $\R^{d+1}$.
Points on the hypersphere can be parametrized using 
hyperspherical coordinate systems.
Any parametrization of the hypersphere $\Si^d$, must have $(\bfx,\bfx)=x_0^2+\ldots+x_d^2=R^2$,
with $R>0$.  The distance between two points on the hypersphere is given by
\begin{equation}
d(\bfx,\bfxp)=R\gamma
=R\cos^{-1}\left(\frac{(\bfx,\bfxp)}{(\bfx,\bfx)(\bfxp,\bfxp)} \right)
=R\cos^{-1}\left(\frac{1}{R^2}(\bfx,\bfxp)\right).
\label{cosgamma}
\end{equation}
This is evident from the fact that the geodesics on $\Si^d$ are great circles 
(i.e.~intersections of $\Si^d$ with planes through the origin) with constant speed 
parametrizations (see p.~82 in Lee (1997) \cite{Lee}).

Accordingly, we now look at the geodesic distance
function on the $d$-dimensional positive radius pseudo-sphere $\Hi_R^d.$ 
Distances between two points on the positive radius pseudo-sphere
are given by the hyperangle between two arbitrary vectors, from the origin, 
in the ambient Minkowski space.  Any parametrization of the hyperboloid $\Hi_R^d$, must 
have $[\bfx,\bfx]=R^2$.  The geodesic distance $\rho\in[0,\infty)$ between two points 
$\bfx,\bfxp\in\Hi_R^d$ is given by
\begin{equation}
d(\bfx,\bfxp)=R\cosh^{-1}\left(\frac{[\bfx,\bfxp]}{[\bfx,\bfx][\bfxp,\bfxp]} \right)
=R\cosh^{-1}\left(\frac{1}{R^2}[\bfx,\bfxp]\right),
\label{dhyperboloid}
\end{equation}
where the inverse hyperbolic cosine with argument $x\in(1,\infty)$ is 
given by (see (4.37.19) in Olver {\it et al.} (2010) \cite{NIST})
$\cosh^{-1}x=\log\left(x+\sqrt{x^2-1}\right).$
Geodesics on $\Hi_R^d$ are great hyperbolas 
(i.e.~intersections of $\Hi_R^d$ with planes through the origin) with constant speed 
parametrizations (see p.~84 in Lee (1997) \cite{Lee}).
We also define a global function $\rho:\Hi^d\times\Hi^d\to[0,\infty)$
which represents the projection of global geodesic distance function
(\ref{dhyperboloid})
on $\Hi_R^d$ onto the corresponding unit radius hyperboloid $\Hi^d$, namely
\begin{equation}
\rho(\wbfx,\wbfxp):=d(\bfx,\bfxp)/R,
\label{rhodefn}
\end{equation}
where $\wbfx=\bfx/R$ and $\wbfxp=\bfxp/R$.  Note that when we refer to $d(\wbfx,\wbfxp)$ below,
we specifically mean that projected distance given by (\ref{rhodefn}).

\subsection{Laplace's equation and harmonics on the hyperboloid}
\label{Laplaceequationandharmonicsonthehyperboloid}

Parametrizations of a submanifold embedded in either a Euclidean or Minkowski space
is given in terms of coordinate systems whose coordinates are curvilinear.  These are 
coordinates based on some transformation that converts the standard Cartesian 
coordinates in the ambient space to a coordinate system with the same number of 
coordinates as the dimension of the submanifold in which the coordinate lines are curved. 

The Laplace-Beltrami operator (Laplacian) in curvilinear coordinates
${\mathbf{\xi}}=(\xi^1,\ldots,\xi^d)$ on a $d$-dimensional Riemannian manifold 
(a manifold together with a Riemannian metric $g$) is given by
\begin{equation}
\Delta=\sum_{i,j=1}^d\frac{1}{\sqrt{|g|}}
\frac{\partial}{\partial \xi^i}
\left(\sqrt{|g|}g^{ij} 
\frac{\partial}{\partial \xi^j}
 \right),
\label{laplacebeltrami}
\end{equation}

\noindent where $|g|=|\det(g_{ij})|,$ the infinitesimal distance is given by
\begin{equation}
ds^2=\sum_{i,j=1}^{d}g_{ij}d\xi^id\xi^j,\ 
\label{metric}
\end{equation}

\noindent and
\[
\sum_{i=1}^{d}g_{ki}g^{ij}=\delta_k^j,
\]
where $\delta_i^j\in\{0,1\}$ with $i,j\in\Z$, is the Kronecker delta defined such that
\begin{equation}
\delta_i^j:=
\left\{ \begin{array}{ll}
\displaystyle 1 &\qquad\mathrm{if}\ i=j, \nonumber \\[0.1cm]
\displaystyle 0 &\qquad\mathrm{if}\ i\ne j.
\end{array} \right.
\label{Kronecker}
\end{equation}
For a submanifold, the relation between the metric 
tensor in the ambient space 
and $g_{ij}$ of (\ref{laplacebeltrami}) and (\ref{metric}) is
\[
g_{ij}({\mathbf{\xi}})=\sum_{k,l=0}^dG_{kl}\frac{\partial x^k}{\partial \xi^i}
\frac{\partial x^l}{\partial \xi^j}.
\]
The ambient space for the hyperboloid is Minkowski, and 
therefore $G_{ij}=\mathrm{diag}(1,-1,\ldots,-1)$.

The set of all geodesic polar coordinate systems corresponds 
to the many ways one can put coordinates on a hyperbolic hypersphere, i.e.,
the Riemannian submanifold $U\subset\Hi_R^d$ defined for a fixed $\bfxp\in\Hi_R^d$
such that $d(\bfx,\bfxp)=b=const,$ where $b\in(0,\infty)$.
These are coordinate systems which correspond to subgroup chains
starting with $O(d,1) \supset O(d) \supset \cdots$,
with standard geodesic polar coordinates given by (\ref{standardhyp}) 
being only one of them. (For a thorough description of these see section  X.5 in 
Vilenkin (1968) \cite{Vilen}.)
They all share the property that they are described by 
$(d+1)$-variables: $r\in[0,\infty)$ plus $d$-angles each being given by the 
values $[0,2\pi)$, $[0,\pi]$, $[-\pi/2,\pi/2]$ or $[0,\pi/2]$ 
(see Izmest'ev {\it et al.}~(1999, 2001) \cite{IPSWa, IPSWb}). 

In any of the geodesic polar coordinate systems, the 
geodesic distance between two points on the submanifold
is given by (cf.~(\ref{dhyperboloid}))
\begin{equation}
d(\bfx,\bfxp)
=R\cosh^{-1}
\bigl( \cosh r\cosh r^\prime - \sinh r\sinh r^\prime\cos\gamma\bigr),
\label{diststandard}
\end{equation}
where $\gamma$ is the unique separation angle given in each hyperspherical coordinate system.
For instance, the separation angle in standard geodesic polar coordinates
(\ref{standardhyp})
is given by the formula
\begin{equation}
\fl\displaystyle\cos\,\gamma=\cos(\phi-\phi^\prime)
\prod_{i=1}^{d-2}\sin\theta_i{\sin\theta_i}^\prime
+\sum_{i=1}^{d-2}\cos\theta_i{\cos\theta_i}^\prime
\prod_{j=1}^{i-1}\sin\theta_j{\sin\theta_j}^\prime.
\label{prodform}
\end{equation}
Corresponding separation angle formulae for any geodesic polar
coordinate system can be computed using (\ref{cosgamma}), (\ref{dhyperboloid}),
and the associated formulae for the appropriate inner-products.

The infinitesimal distance in a geodesic polar coordinate system 
on this submanifold is
\begin{equation}
ds^2=R^2(dr^2+\sinh^2r\ d\gamma^2),
\label{stanhypmetric}
\end{equation}
where an appropriate expression for $\gamma$ in a curvilinear coordinate system is given.
If one combines 
(\ref{standardhyp}), 
(\ref{laplacebeltrami}), 
(\ref{prodform})
and (\ref{stanhypmetric}), then in a particular geodesic polar coordinate system, 
Laplace's equation on $\Hi_R^d$ is
\begin{equation}
\Delta f=\frac{1}{R^2}\left[\frac{\partial^2f}{\partial r^2}
+(d-1)\coth r\frac{\partial f}{\partial r}
+\frac{1}{\sinh^2r} \Delta_{\Si^{d-1}}f\right]=0,
\label{genhyplap}
\end{equation}
where $\Delta_{\Si^{d-1}}$ is the corresponding Laplace-Beltrami operator
on $\Si^{d-1}$ with unit radius. 

From this point onwards, $\Si^{d-1}$ will always
refer to the $(d-1)$-dimensional unit hypersphere, which is a compact Riemannian 
submanifold with positive constant sectional curvature, embedded in $\R^d$ and
given by the variety $x_1^2+\ldots+x_d^2=1$.

Geodesic polar coordinate systems 
partition $\Hi_R^d$ into a family of concentric $(d-1)$-dimensional 
hyperspheres, each with a radius $r\in(0,\infty),$ on which all possible 
hyperspherical coordinate systems
for $\Si^{d-1}$ may be used
(see for instance, in Vilenkin (1968) \cite{Vilen}).  One then must also 
consider the limiting case for $r=0$ to fill out all of $\Hi_R^d$.
In standard geodesic polar coordinates one can compute the 
normalized hyperspherical harmonics in this space by solving the Laplace equation
using separation of variables which results in a general procedure 
which is given explicitly in Izmest'ev {\it et al.}~(1999, 2001) \cite{IPSWa, IPSWb}. 
These angular harmonics are given as general expressions involving trigonometric functions, 
Gegenbauer polynomials and Jacobi polynomials.

The harmonics in geodesic polar coordinate systems
are given in terms of a radial solution multiplied by the 
angular harmonics.  The angular harmonics are eigenfunctions of the Laplace-Beltrami operator
on $\Si^{d-1}$ with unit radius which satisfy the following eigenvalue problem
\begin{equation}
\Delta_{\Si^{d-1}}Y_l^K(\wbfx)
=-l(l+d-2)Y_l^K(\wbfx),
\label{eq4}
\end{equation}
where 
$\wbfx\in\Si^{d-1}$,
$Y_l^K(\wbfx)$
are normalized hyperspherical harmonics,
$l\in\N_0$ is the 
angular momentum quantum number, and 
$K$ stands for the set of $(d-2)$-quantum numbers identifying 
degenerate harmonics for each $l$. 
The degeneracy as a function of the dimension $d$ tells you
how many linearly independent solutions exist for a particular $l$ value.
The hyperspherical harmonics are normalized such that
\[
\int_{\Si^{d-1}} Y_l^K
(\wbfx)
%(\theta_1,\ldots,\theta_{d-1})
\overline{Y_{l^\prime}^{K^\prime}
%(\theta_1^\prime,\ldots,\theta_{d-1}^\prime)
(\wbfx)
}
d\omega=
\delta_{l}^{l^\prime}
\delta_{K}^{K^\prime},
\]
where $d\omega$ is a volume measure on $\Si^{d-1}$ which is invariant under the isometry group $SO(d)$ 
(cf.~(\ref{eucsphmeasureinv})), and for $x+iy=z\in\C$, $\overline{z}=x-iy$, represents complex conjugation.
The generalized Kronecker delta $\delta_K^{K^\prime}$ 
(cf.~(\ref{Kronecker}))
is defined such that it equals 1 if 
all of the $(d-2)$-quantum numbers identifying degenerate harmonics for each $l$ coincide, 
and equals zero otherwise.

Since the angular solutions (hyperspherical harmonics) are well-known (see 
Chapter IX in Vilenkin (1968) \cite{Vilen}; 
Chapter 11 in Erd{\'e}lyi {\it et al.} (1981) \cite{ErdelyiHTFII}), 
we will now focus on the radial solutions, which 
satisfy the following ordinary differential equation
\[
\frac{d^2u}{dr^2}+(d-1)\coth r\frac{du}{dr}-\frac{l(l+d-2)}{\sinh^2r}u=0.
\]

\noindent Four solutions to this ordinary differential equation 
$u_{1\pm}^{d,l},u_{2\pm}^{d,l}:(1\infty)\to\C$ are given by
\[
{\displaystyle u_{1\pm}^{d,l}(\cosh r)=\frac{1}{\sinh^{d/2-1}r}P_{d/2-1}^{\pm(d/2-1+l)}
(\cosh r)},
\]
\noindent and
\[
{\displaystyle u_{2\pm}^{d,l}(\cosh r)=\frac{1}{\sinh^{d/2-1}r}Q_{d/2-1}^{\pm(d/2-1+l)}
(\cosh r)},
\]
where $P_\nu^\mu,Q_\nu^\mu:(1,\infty)\to\C$ are associated Legendre functions of the first and
second kind respectively (see for instance Chapter 14 in Olver {\it et al.} (2010) \cite{NIST}).

\subsection{Fundamental solution of Laplace's equation on the hyperboloid}

Due to the fact that the space $\Hi_R^d$ is homogeneous with respect to 
its isometry group, the pseudo-orthogonal group $SO(d,1)$, and therefore 
an isotropic manifold, we expect that there exist a fundamental solution
on this space with spherically symmetric dependence.  We specifically expect
these solutions to be given in terms of associated Legendre functions of the 
second kind with argument given by $\cosh r$.  This associated Legendre 
function naturally fits our requirements because it is singular at $r=0$ and 
vanishes at infinity, whereas the associated Legendre functions of the 
first kind, with the same argument, are regular at $r=0$ and singular at infinity.  

In computing a fundamental solution of the Laplacian on $\Hi_R^d$, we know that
\begin{equation}
-\Delta \mch_R^d(\bfx,\bfxp) = \delta_g(\bfx,\bfxp),
\label{eq3}
\end{equation}
where $g$ is the Riemannian metric on $\Hi_R^d$ and 
$\delta_g(\bfx,\bfxp)$ is the Dirac delta function on the manifold $\Hi_R^d$.  
The Dirac delta function
is defined for an open set $U\subset\Hi_R^d$ with $\bfx,\bfxp\in\Hi_R^d$ such that 
\[
\int_U\delta_g(\bfx,\bfxp) d\mbox{vol}_g =
\left\{ \begin{array}{ll}
\displaystyle 1 &\qquad\mathrm{if}\ \bfxp\in U, \nonumber \\[0.1cm]
\displaystyle 0 &\qquad\mathrm{if}\ \bfxp\notin U,
\end{array} \right.
\]
where $d\mbox{vol}_g$ is a volume measure, invariant under the isometry group $SO(d,1)$ of 
the Riemannian manifold $\Hi_R^d$, given (in standard geodesic polar coordinates) by 
\begin{equation}
\fl d\mbox{vol}_g=
R^d\sinh^{d-1}rd\omega:=
R^d\sinh^{d-1}r
\sin^{d-2}\theta_{d-1}\cdots\sin\theta_2 d\theta_{1}\cdots d\theta_{d-1}.
\label{eucsphmeasureinv}
\end{equation}
Notice that as $r\to 0^+$ that $d\mbox{vol}_g$ goes to the Euclidean measure, invariant under the Euclidean 
motion group $E(d)$, in spherical coordinates.  Therefore
in spherical coordinates, we have the following
\begin{equation}
\delta_g(\bfx,\bfxp)=\frac{\delta(r-r^\prime)}{R^d\sinh^{d-1}r^\prime}
\frac{\delta(\theta_1-\theta_1^\prime)\cdots\delta(\theta_{d-1}-\theta_{d-1}^\prime)}
{\sin\theta_2^\prime\cdots\sin^{d-2}\theta_{d-1}^\prime}.
\label{diracdeltasubg}
\end{equation}

In general since we can add any harmonic function to a fundamental solution 
for the Laplacian and 
still have a fundamental solution, we will use this freedom to make our 
fundamental solution as simple as possible.  
It is reasonable to expect that there exists a particular spherically
symmetric fundamental solution
$\mch_R^d(\bfx,\bfxp)$
on the hyperboloid
with pure radial $\rho(\wbfx,\wbfxp)=d(\bfx,\bfxp)/R$
(cf.~(\ref{rhodefn})) and constant angular dependence
(invariant under rotations centered about the origin),
due to
the influence of the point-like nature of the Dirac delta function.
For a spherically symmetric solution to the Laplace
equation,
the
corresponding $\Delta_{\Si^{d-1}}$
term
vanishes since only the $l=0$ term survives.
In other words, we expect there to exist a fundamental solution of Laplace's
equation such that
$\mch_R^d(\bfx,\bfxp)=f(\rho)$.
\medskip

In Cohl \& Kalnins (2011) \cite{CohlKalI},
we have proven that on the $R$-radius hyperboloid $\Hi_R^d$, a fundamental solution 
of Laplace's equation can be given as follows.
%\newpage
\begin{thm}
Let $d\in\{2,3,\ldots\}.$  Define $\mcI_d:(0,\infty)\to\R$ as
\[
\mcI_d(\rho):=\int_\rho^\infty\frac{dx}{\sinh^{d-1}x},
\]
$\bfx,\bfxp\in\Hi_R^d$, and
$\mch_R^d:(\Hi_R^d\times\Hi_R^d)\setminus\{(\bfx,\bfx):\bfx\in\Hi_R^d\}\to\R$
defined such that
\[
\mch_R^d({\bf x},{\bf x}^\prime):=
{\displaystyle \frac{\Gamma\left(d/2\right)}{2\pi^{d/2}R^{d-2}}\mcI_d(\rho)},
\label{thmh1d}
\]
where $\rho:=\cosh^{-1}\left([\wbfx,\wbfxp]\right)$ is the geodesic distance
between $\wbfx$ and $\wbfxp$ on the pseudo-sphere of unit radius $\Hi^d$,
with
$\wbfx=\bfx/R,$
$\wbfxp=\bfxp/R$,
then $\mch_R^d$ is a fundamental solution for $-\Delta$
where $\Delta$ is the
Laplace-Beltrami operator on $\Hi_R^d$.
Moreover,\\[-0.3cm]
\begin{eqnarray*}
\fl\mcI_d(\rho)=
\left\{ \begin{array}{ll}
\displaystyle (-1)^{d/2-1}\frac{(d-3)!!}{(d-2)!!}
\Biggl[\log\coth \frac{\rho}{2}
+\cosh \rho\sum_{k=1}^{d/2-1}\frac{(2k-2)!!(-1)^k}{(2k-1)!!\sinh^{2k}\rho}\Biggr]
&\hspace{-0.2cm}\mathrm{if}\  d\ \mathrm{even}, \\[0.6cm]
\left\{ \begin{array}{l}
\displaystyle
(-1)^{(d-1)/2}\Biggl[\frac{(d-3)!!}{(d-2)!!}\\[0.4cm]
\displaystyle \hspace{1.2cm}+\left(\frac{d-3}{2}\right)!
\sum_{k=1}^{(d-1)/2}
\frac{(-1)^k\coth^{2k-1}\rho}
{(2k-1)(k-1)!((d-2k-1)/2)!}\Biggr],
\\[0.45cm]
\mathrm{or} \\[0.0cm]
\displaystyle
(-1)^{(d-1)/2}\frac{(d-3)!!}{(d-2)!!}\left[1+\cosh\rho
\sum_{k=1}^{(d-1)/2}
\frac{(2k-3)!!(-1)^k}{(2k-2)!!\sinh^{2k-1}\rho}\right],
\end{array} \right\}  &\hspace{-0.2cm}\mathrm{if}\  d\ \mathrm{odd}.
\end{array} \right.\\[0.3cm]
\hspace{-1.47cm}
=\frac{1}{(d-1)\cosh^{d-1}\rho}\,
{}_2F_1\left(\frac{d-1}{2},\frac{d}{2};\frac{d+1}{2};\frac{1}{\cosh^2\rho}\right),\\[0.3cm]
\hspace{-1.47cm}
=\frac{1}{(d-1)\cosh \rho\,\sinh^{d-2}\rho}\,
{}_2F_1\left(\frac12,1;\frac{d+1}{2};\frac{1}{\cosh^2\rho}\right),\\[0.3cm]
\hspace{-1.47cm}
%=\frac{e^{-i\pi(d/2-1)}}{2^{d/2-1}\Gamma\left(\frac{d}{2}\right)\sinh^{d/2-1}\rho}\,
=\frac{e^{-i\pi(d/2-1)}}{2^{d/2-1}\Gamma\left(d/2\right)\sinh^{d/2-1}\rho}\,
Q_{d/2-1}^{d/2-1}(\cosh\rho),\\[0.3cm]
\end{eqnarray*}\\[-1.0cm]
where $!!$ is the double factorial, ${}_2F_1$ is the Gauss hypergeometric function,
and $Q_\nu^\mu$ is the associated Legendre function of the second kind.
\end{thm}

\medskip
For a proof of this theorem, see Cohl \& Kalnins (2011) \cite{CohlKalI}.

\section{Fourier expansions for a Green's function on the hyperboloid}
\label{FourierexpansionsforaGreensfunctioninthehyperboloidmodel}

Now we compute the Fourier expansions for 
a fundamental solution of the Laplace-Beltrami operator on $\Hi_R^d$.

\subsection{Fourier expansion for a fundamental solution of the Laplacian on $\Hi_R^2$}
\label{FourierexpansionforaGreensfunctioninHi2}

The generating function for Chebyshev polynomials
of the first kind (Fox \& Parker (1968) \cite{FoxParker}, p.~51) is given as
\begin{equation}
\frac{1-z^2}{1+z^2-2xz}=\sum_{n=0}^\infty \epsilon_n T_n(x) z^n,
\label{genfunchebyshevfirst}
\end{equation}
where $|z|<1,$ $T_{n}:[-1,1]\to\R$ is the Chebyshev polynomial of the first kind
defined as $T_l(x):=\cos(l\cos^{-1}x),$ 
and $\epsilon_n:=2-\delta_{n}^{0}$ is the Neumann factor 
(see p.~744 in Morse \& Feshbach (1953) \cite{MorseFesh}),
commonly-occurring in Fourier cosine series.
If substitute $z=e^{-\eta}$ with $\eta\in(0,\infty)$ in
(\ref{genfunchebyshevfirst}), then we obtain
\begin{equation}
\frac{\sinh\eta}{\cosh\eta-\cos\psi}=
\sum_{n=0}^\infty \epsilon_n\cos(n\psi)e^{-n\eta}.
\label{genfuncheb}
\end{equation}
Integrating both sides of (\ref{genfuncheb}) with respect to $\eta$, we obtain the following 
formula (cf.~Magnus, Oberhettinger \& Soni (1966) \cite{MOS}, p.~259)
\begin{equation}
\log\left( 
1+z^2-2z\cos\psi
\right)=-
2\sum_{n=1}^\infty
\frac{\cos(n\psi)}{n}z^n.
\label{logzequation}
\end{equation}

In Euclidean space $\R^d$, a Green's function for Laplace's 
equation (fundamental solution for the Laplacian) 
is well-known and is given in the following theorem
(see Folland (1976) \cite{Fol3}; p.~94, Gilbarg \& Trudinger (1983) \cite{GT}; 
p.~17, Bers {\it et al.}~(1964) \cite{BJS}, p.~211).
\begin{thm}
Let $d\in\N$. Define
\[
\mcg^d({\bf x},{\bf x}^\prime)=
\left\{ \begin{array}{ll}
\displaystyle\frac{\Gamma(d/2)}{2\pi^{d/2}(d-2)}\|{\bf x}-{\bf x}^\prime\|^{2-d}
& \qquad\mathrm{if}\ d=1\mathrm{\ or\ }d\ge 3,\\[10pt]
\displaystyle\frac{1}{2\pi}\log\|{\bf x}-{\bf x}^\prime\|^{-1}
& \qquad\mathrm{if}\  d=2, \nonumber
\end{array} \right. 
\]
then $\mcg^d$ is a fundamental solution 
for $-\Delta$
in Euclidean space $\R^d$,
where $\Delta$ is the Laplace operator in $\R^d$.
\label{thmg1n}
\end{thm}
Therefore if we take $z=r_</r_>$ in (\ref{logzequation}), 
where $r_\lessgtr:={\min \atop \max}\{r,r^\prime\}$ with $r,r^\prime\in[0,\infty),$
then using polar coordinates, we can derive 
the Fourier expansion for a fundamental solution of the Laplacian in Euclidean 
space for $d=2$ (cf.~Theorem \ref{thmg1n}), namely
\begin{equation}
\g^2:=\log \|\bfx-\bfxp\| =\log r_> 
- \sum_{n=1}^\infty \frac{\cos(n(\phi-\phi^\prime))}{n}
\biggl(\frac{r_<}{r_>} \biggr)^n,
\label{eucfundsolfourexpansd2}
\end{equation}
where 
$\g^2=-2\pi\mcg^2$
(cf.~Theorem \ref{thmg1n}).  On the hyperboloid for $d=2$ we have a fundamental 
solution of Laplace's equation given by 
\[
\h^2:=\log\coth\frac12 d(\wbfx,\wbfxp)=\frac12\log
\frac{\cosh d(\wbfx,\wbfxp)+1}
{\cosh d(\wbfx,\wbfxp)-1},
\]
where
$\h^2=2\pi\mch_R^2$ (cf.~Theorem \ref{thmh1d} 
and (\ref{backhd}) below).
Note that because of the $R^{d-2}$ dependence of a fundamental solution of
Laplace's equation for $d=2$ in Theorem \ref{thmh1d}, there is no strict
dependence on $R$ for $\mch_R^2$ or $\h^2$, but will retain the notation nonetheless.
In standard geodesic polar coordinates on $\Hi_R^2$ 
(cf.~(\ref{standardhyp})), 
using 
(\ref{diststandard}) 
and
$\cos\gamma=\cos(\phi-\phi^\prime)$
(cf.~(\ref{prodform}))
produces
\[
\cosh d(\wbfx,\wbfxp)=\cosh r\cosh r^\prime-\sinh r\sinh r^\prime\cos(\phi-\phi^\prime),
\]
\noindent therefore
\[
\h^2=\frac12\log
\frac{\cosh r\cosh r^\prime+1-\sinh r\sinh r^\prime\cos(\phi-\phi^\prime)}
{\cosh r\cosh r^\prime-1-\sinh r\sinh r^\prime\cos(\phi-\phi^\prime)}.
\]
Replacing $\psi=\phi-\phi^\prime$ and rearranging the logarithms yield
\[
\h^2=\frac12\log
\frac{\cosh r\cosh r^\prime+1}
{\cosh r\cosh r^\prime-1}
+\frac12\log\left(1-
%\frac{\sinh r\sinh r^\prime}{\cosh r\cosh r^\prime-1}
z_+
\cos\psi \right)
-\frac12\log\left(1-
%\frac{\sinh r\sinh r^\prime}{\cosh r\cosh r^\prime+1}
z_-
\cos\psi \right),
\]
where 
\[
z_\pm:=
\frac{\sinh r\sinh r^\prime}{\cosh r\cosh r^\prime\pm 1}.
\]
Note that $z_\pm\in(0,1)$ for $r,r^\prime\in(0,\infty)$.
We have the following MacLaurin series 
\[
\log(1-x)=-\sum_{n=1}^\infty \frac{x^n}{n},
\]
where $x\in[-1,1)$.  Therefore away from the singularity at $\bfx=\bfxp$ we have
\begin{equation}
\lambda_\pm:=\log\left(1-
%\frac{\sinh r\sinh r^\prime}{\cosh r\cosh r^\prime\pm 1}
z_\pm
\cos\psi \right)=
-\sum_{k=1}^\infty\frac{z_\pm^k}{k}
%\left[ 
%\frac{\sinh r\sinh r^\prime}{\cosh r\cosh r^\prime\pm 1}
%\right]^k 
\cos^k\psi.
\label{loghyp}
\end{equation}
We can expand the powers of cosine using the following trigonometric identity
\[
\cos^k\psi=\frac{1}{2^k}\sum_{n=0}^k
%\binom{k}{n} 
\left( \begin{array}{c}
\displaystyle \!\!k\\[1pt]
\displaystyle \!\!n
\end{array} \!\! \right)
\cos[(2n-k)\psi],
\]

\noindent which is the standard expansion for powers using 
Chebyshev polynomials
(see for instance
p.~52 in Fox \& Parker (1968) \cite{FoxParker}).  Inserting this expression in
(\ref{loghyp}), we obtain the following double-summation expression 

\begin{equation}
%\log\left(1-\frac{\sinh r\sinh r^\prime}
%{\cosh r\cosh r^\prime\pm 1}\cos\psi \right)=
\lambda_\pm=
-\sum_{k=1}^\infty
\sum_{n=0}^k
\frac{z_\pm^k}{2^kk}
%\left[ 
%\frac{\sinh \sinh r^\prime}{\cosh r\cosh r^\prime\pm 1}
%\right]^k 
%\binom{k}{n}
\left( \begin{array}{c}
\displaystyle \!\!k\\[1pt]
\displaystyle \!\!n
\end{array} \!\! \right)
\cos[(2n-k)\psi].
\label{dbsumhyp}
\end{equation}

Now we perform a double-index replacement in (\ref{dbsumhyp}).
We break this
sum into two separate sums, one for $k\le 2n$ and another for $k\ge 2n$.  There 
is an overlap when both sums satisfy the equality, and in that situation we must 
halve after we sum over both sums.  
If $k\le 2n$, make the substitution 
$k^\prime=k-n$ and $n^\prime=2n-k$.  It follows
that $k=2k^\prime+n^\prime$ and $n=n^\prime+k^\prime$, therefore
\[
%\binom{k}{n}
\left( \begin{array}{c}
\displaystyle \!\!k\\[1pt]
\displaystyle \!\!n
\end{array} \!\! \right)
=
%\binom{2k^\prime+n^\prime}{n^\prime+k^\prime}
\left( \begin{array}{c}
\displaystyle \!\!2k^\prime+n^\prime\\[1pt]
\displaystyle \!\!n^\prime+k^\prime
\end{array} \!\! \right)
=
%\binom{2k^\prime+n^\prime}{k^\prime}
\left( \begin{array}{c}
\displaystyle \!\!2k^\prime+n^\prime\\[1pt]
\displaystyle \!\!n^\prime+k^\prime
\end{array} \!\! \right)
.
\]

\noindent If $k\ge 2n$ make the substitution $k^\prime=n$ and $n^\prime=k-2n$.  
Then 
$k= 2k^\prime + n^\prime$ and $n=k^\prime$, therefore
\[
%\binom{k}{n}
\left( \begin{array}{c}
\displaystyle \!\!k\\[1pt]
\displaystyle \!\!n
\end{array} \!\! \right)
=
%\binom{2k^\prime+ n^\prime}{k^\prime}
\left( \begin{array}{c}
\displaystyle \!\!2k^\prime+n^\prime\\[1pt]
\displaystyle \!\!n
\end{array} \!\! \right)
=
%\binom{2k^\prime+n^\prime}{k^\prime+n^\prime}
\left( \begin{array}{c}
\displaystyle \!\!2k^\prime+n^\prime\\[1pt]
\displaystyle \!\!k^\prime+n^\prime
\end{array} \!\! \right)
,
\]

\noindent where the equalities of the binomial coefficients
are confirmed using
the following identity
\begin{equation*}
%\binom{n}{k}
\left( \begin{array}{c}
\displaystyle \!\!n\\[1pt]
\displaystyle \!\!k
\end{array} \!\! \right)
=
%\binom{n}{n-k}
\left( \begin{array}{c}
\displaystyle \!\!n\\[1pt]
\displaystyle \!\!n-k
\end{array} \!\! \right)
,
\end{equation*}
where $n,k\in\Z$, except where $k<0$ or $n-k<0$.
To take into account the double-counting which occurs when $k=2n$
(which occurs when $n^\prime=0$), we introduce a factor of $\epsilon_{n^\prime}/2$ 
into the expression
(and relabel $k^\prime\mapsto k$ and $n^\prime\mapsto n$).  We are left with
\begin{equation}
\fl\lambda_\pm=-\frac12\sum_{k=1}^\infty
\frac{z_\pm^{2k}}{2^kk}
%\binom{2k}{k}
\left( \begin{array}{c}
\displaystyle \!\!2k\\[1pt]
\displaystyle \!\!k
\end{array} \!\! \right)
-2\sum_{n=1}^\infty\cos(n\psi)\sum_{k=0}^\infty
\frac{z_\pm^{2k+n}}{2^{2k+n}(2k+n)}
%\binom{2k+n}{k}
\left( \begin{array}{c}
\displaystyle \!\!2k+n\\[1pt]
\displaystyle \!\!k
\end{array} \!\! \right)
.
\label{lambdapmresult}
\end{equation}
If we substitute
\[
%\binom{2k}{k}
\left( \begin{array}{c}
\displaystyle \!\!2k\\[1pt]
\displaystyle \!\!k
\end{array} \!\! \right)
=\frac{2^{2k}\left(\frac12\right)_k}{k!}
\]
into the first term of (\ref{lambdapmresult}), then we obtain
\[
\fl I_\pm:=-\frac12\sum_{k=1}^\infty
\frac{
\left(\frac12\right)_k
z_\pm^{2k}
}{k!k}=
-\int_0^{z_\pm}\frac{dz_\pm^\prime}{z_\pm^\prime}
\sum_{k=1}^\infty
\frac{\left(\frac12\right)_k {z_\pm^\prime}^{2k}}{k!}
=-\int_0^{z_\pm} \frac{d{z_\pm^\prime}}{z_\pm^\prime}
\left[
\frac{1}{\sqrt{1-{z_\pm^\prime}^2}}-1
\right].
\]
We are left with
\[
\fl I_\pm=
-\log 2+\log\left(1+\sqrt{1-z_\pm^2} \right)
=
-\log 2+\log\left(
\frac{(\cosh r_>\pm 1)(\cosh r_<+1)}{\cosh r\cosh r^\prime\pm 1}
\right).
\]
If we substitute
\[
\displaystyle 
%\binom{2k+n}{k}
\left( \begin{array}{c}
\displaystyle \!\!2k+n\\[1pt]
\displaystyle \!\!k
\end{array} \!\! \right)
=\frac{\displaystyle 2^{2k}
\left(\frac{n+1}{2}\right)_k
\left(\frac{n+2}{2}\right)_k
}{k!(n+1)_k},
\]
into the second term of (\ref{lambdapmresult}), then the Fourier coefficient reduces to
\begin{eqnarray}
J_\pm&:=&\frac{1}{2^{n-1}}\sum_{k=0}^\infty
\frac{\displaystyle 
\left(\frac{n+1}{2}\right)_k
\left(\frac{n+2}{2}\right)_k
}{\displaystyle k!(n+1)_k}
\frac{z_\pm^{2k+n}}{2k+n}\nonumber\\[0.2cm]
&=&
\frac{1}{2^{n-1}}\int_0^{z_\pm}
dz_\pm^\prime {z_\pm^\prime}^{n-1}
\sum_{k=0}^\infty
\frac{\displaystyle
\left(\frac{n+1}{2}\right)_k
\left(\frac{n+2}{2}\right)_k
}{k!(n+1)_k}
{z_\pm^\prime}^{2k}.\nonumber
\end{eqnarray}
The series in the integrand is a Gauss hypergeometric 
function
which can be given as
\[
\sum_{k=0}^\infty
\frac{\displaystyle
\left(\frac{n+1}{2}\right)_k
\left(\frac{n+2}{2}\right)_k
}{k!(n+1)_k}
z^{2k}=\frac{2^nn!}{z^n\sqrt{1-z^2}}P_0^{-n}\left(\sqrt{1-z^2}\right),
\]
where $P_0^{-n}$ is an associated 
Legendre function
of the first kind 
with vanishing degree and order given by $-n$.  This is a consequence of
\[
\fl {}_2F_1\left(a,b;a+b-\frac12;x \right)=
2^{2+b-3/2}\Gamma\left(a+b-\frac12\right)\frac{x^{(3-2a-2b)/4}}{\sqrt{1-x}}
P_{b-a-1/2}^{3/2-a-b}\left(\sqrt{1-x} \right),
\]
where $x\in(0,1)$ (see for instance Magnus, Oberhettinger \& Soni (1966) \cite{MOS}, p.~53),
and the Legendre function is evaluated using 
(cf.~(8.1.2) in Abramowitz \& Stegun (1972) \cite{Abra})
\[
P_0^{-n}(x)=\frac{1}{n!}\left(\frac{1-x}{1+x} \right)^{n/2},
\]
where $n\in\N_0$.  
Therefore the Fourier coefficient is given by
\[
J_\pm=2\int_{\sqrt{1-z_\pm^2}}^1
\frac{dz_\pm^\prime}{1-{z_\pm^\prime}^2}\left(
\frac{1-z_\pm^\prime}{1+z_\pm^\prime}
\right)^{n/2}=
\frac{2}{n}\left[ \frac{1-\sqrt{1-z_\pm^2}}{1+\sqrt{1-z_\pm^2}}\right]^{n/2}.
\]
Finally we have
\begin{eqnarray*}
\lambda_\pm=
-\log 2
+\log\left(\frac{(\cosh r_>\pm 1)(\cosh r_<+1)}{\cosh r\cosh r^\prime\pm 1}
\right)\nonumber\\[0.0cm]
\hspace{1.3cm}-2\sum_{n=1}^\infty \frac{\cos(n\psi)}{n}
\left[
\frac
{(\cosh r_>\mp 1)(\cosh r_<-1)}
{(\cosh r_>\pm 1)(\cosh r_<+1)}
\right]^{n/2},\nonumber
\end{eqnarray*}
and the Fourier expansion for a fundamental solution of Laplace's equation 
for the $d=2$ hyperboloid is given by
\begin{eqnarray}
\fl \h^2=\frac12\log\frac{\cosh r_>+1}{\cosh r_>-1}\nonumber\\[0.0cm]
\hspace{-2.2cm}{}\displaystyle+\sum_{n=1}^\infty
\frac{\cos(n(\phi-\phi^\prime))}{n}
\left[ \frac{\cosh r_<-1}{\cosh r_<+1}\right]^{n/2}
\left\{
\left[
\frac{\cosh r_>+1}{\cosh r_>-1}
\right]^{n/2}
-\left[
\frac{\cosh r_>-1}{\cosh r_>+1}
\right]^{n/2}
\right\}.
\label{expansiondequals2}
\end{eqnarray}
This exactly matches up to the Euclidean Fourier 
expansion $\g^2$ (\ref{eucfundsolfourexpansd2}) as $r,r^\prime\to 0^+$.

\subsection{Fourier expansion for a fundamental solution of the Laplacian on $\Hi_R^3$}
\label{FourierexpansionforaGreensfunctioninHi3}

The Fourier expansion for a fundamental solution of the Laplacian in three-dimensional 
Euclidean space (here given in standard spherical coordinates
$\bfx=(r\sin\theta\cos\phi,r\sin\theta\sin\phi,r\cos\theta)$) is given by 
(cf.~Theorem \ref{thmg1n}, and see (1.3) in Cohl {\it et al.} (2001) \cite{CRTB})
\begin{eqnarray*}
\fl \mcg^3\simeq \g^3:=\frac{1}{\|\bfx-\bfxp\|}\nonumber\\[0.0cm]
\hspace{-0.9cm}=\frac{1}{\pi\sqrt{rr^\prime\sin\theta\sin\theta^\prime}}
\sum_{m=-\infty}^\infty e^{im(\phi-\phi^\prime)} Q_{m-1/2}\left(
\frac{r^2+{r^\prime}^2-2rr^\prime\cos\theta\cos\theta^\prime}
{2rr^\prime\sin\theta\sin\theta^\prime}
\right).
\end{eqnarray*}
These associated Legendre functions, toroidal harmonics,
are given in terms of complete elliptic integrals of the first
and second kind (cf.~(22--26) in Cohl \& Tohline (1999) \cite{CT}).
Since $Q_{-1/2}(z)$ is given through 
(cf.~(8.13.3) in Abramowitz \& Stegun (1972) \cite{Abra})
\[
Q_{-1/2}(z)=
\sqrt{\frac{2}{z+1}}
K\left(\sqrt{\frac{2}{z+1}}\right),
\]
the $m=0$ component for $\g^3$ is given by
\begin{equation}
\fl \left.\g^3\right|_{m=0}=\frac{2}
{\pi\sqrt{r^2+{r^\prime}^2-2rr^\prime\cos(\theta+\theta^\prime)}}
K\left( 
\sqrt{\frac{4rr^\prime\sin\theta\sin\theta^\prime}{r^2+{r^\prime}^2-2rr^\prime\cos(\theta+\theta^\prime)}}
\right).
\label{gm0euc}
\end{equation}
%where $K$ is Legendre's complete elliptic integral of the first kind.

A fundamental solution of the Laplacian in standard geodesic polar coordinates on 
$\Hi_R^3$ is given by
(cf.~Theorem \ref{thmh1d}
and (\ref{backhd}) below).
\begin{eqnarray}
\displaystyle \hspace{-1.85cm}\h^3(\wbfx,\wbfxp):=\coth d(\wbfx,\wbfxp)-1=
\frac{\cosh d(\wbfx,\wbfxp)}
{\sqrt{\cosh^2d(\wbfx,\wbfxp)-1}}-1\nonumber\\[0.0cm]
\hspace{3.3cm}=\frac{\cosh r\cosh r^\prime-\sinh r\sinh r^\prime\cos\gamma}
{\sqrt{(\cosh r\cosh r^\prime-\sinh r\sinh r^\prime\cos\gamma)^2-1}}-1,\nonumber
\end{eqnarray}
where $\h^3=4\pi R\mch_R^3,$ and $\bfx,\bfxp\in\Hi_R^3$,
such that $\wbfx=\bfx/R$ and $\wbfxp=\bfxp/R$.
In standard geodesic polar coordinates 
(cf.~(\ref{prodform})) we have
\begin{equation}
\cos\gamma=\cos\theta\cos\theta^\prime+\sin\theta\sin\theta^\prime\cos(\phi-\phi^\prime).
\label{cosgammah3}
\end{equation}
Replacing $\psi=\phi-\phi^\prime$ and defining
\[
A:=\cosh r\cosh r^\prime-\sinh r\sinh r^\prime\cos\theta\cos\theta^\prime,
\]
and
\[
B:=\sinh r\sinh r^\prime\sin\theta\sin\theta^\prime,
\]
we have in the standard manner, the Fourier coefficients ${\sf H}_m^{1/2}:[0,\infty)^2\times[0,\pi]^2\to\R$ 
of the expansion (cf.~(\ref{fulldFourierexpansion}) below)
\begin{equation}
\h^3(\wbfx,\wbfxp)=\sum_{m=0}^\infty 
%e^{im(\phi-\phi^\prime)}
\cos(m(\phi-\phi^\prime))
{\sf H}_m^{1/2}(r,r^\prime,\theta,\theta^\prime),
\label{azimuthalfourierexpansionh3}
\end{equation}
defined by
\begin{equation}
{\displaystyle {\sf H}_m^{1/2}(r,r^\prime,\theta,\theta^\prime):=-\delta_{n}^0+\frac{\epsilon_m}{\pi}}
{\displaystyle \int_0^\pi\frac{\left(A/B-\cos\psi\right)\cos(m\psi) d\psi}
{\sqrt{
\left(\cos\psi-\frac{A+1}{B}\right)
\left(\cos\psi-\frac{A-1}{B}\right)
}}}.
\label{azimuthalFourierintegralh3}
\end{equation}
If we make the substitution $x=\cos\psi$, this integral can be converted to
\begin{equation}
\fl{\displaystyle {\sf H}_m^{1/2}(r,r^\prime,\theta,\theta^\prime)=-\delta_{n}^0+
\frac{\epsilon_m}{\pi}}
{\displaystyle \int_{-1}^{1}\frac{\left(A/B-x\right)T_m(x)dx}
{\sqrt{
(1-x)(1+x)
\left(x-\frac{A+1}{B}\right)
\left(x-\frac{A-1}{B}\right)
}}},
\label{Amxtm}
\end{equation}
where $T_m$ is the Chebyshev polynomial
of the first kind.
Since $T_m(x)$
is expressible as a finite sum 
over powers of $x$, (\ref{Amxtm}) involves the square root of a quartic 
multiplied by a rational function of $x$, which by definition is
an elliptic integral (see for instance Byrd \& Friedman (1954) \cite{ByrdFriedman}).
We can directly compute 
(\ref{Amxtm})
using Byrd \& Friedman (1954)
(\cite{ByrdFriedman}, (253.11)).  
If we define
\begin{equation}
d:=-1,\ y:=-1,\ c:=1,\ b:=\frac{A-1}{B},\ a:=\frac{A+1}{B},
\label{abcdy}
\end{equation}
(clearly $d\le y<c<b< a$), then we can express the Fourier coefficient
(\ref{Amxtm}), as a linear combination of integrals, each of the form
(see Byrd \& Friedman (1954) \cite{ByrdFriedman}, (253.11))
\begin{equation}
\int_y^c\frac{x^pdx}{\sqrt{(a-x)(b-x)(c-x)(x-d)}}=c^pg\int_0^{u_1}
\left[ 
\frac{1-\alpha_1^2\mathrm{sn}^2u}
{1-\alpha^2\mathrm{sn}^2u}
\right]^pdu,
\label{ourellipint}
\end{equation}
where $p\in\{0,\ldots,m+1\}$.
In this expression $\mathrm{sn}$ is a Jacobi 
elliptic function (see for instance Chapter 22 in
Olver {\it et al.} (2010) \cite{NIST}).

Byrd \& Friedman (1954) \cite{ByrdFriedman} give a procedure for computing 
(\ref{ourellipint}) for all $m\in\N_0$.
These integrals will be given in terms of 
complete elliptic integrals of the first 
three kinds
(see the discussion in Byrd \& Friedman (1954) \cite{ByrdFriedman}, p.~201,~204,~and p.~205).
To this effect, we have the following definitions from (253.11) in Byrd \& Friedman (1954) \cite{ByrdFriedman}, namely
\[
\alpha^2=\frac{c-d}{b-d}<1,
\]
\[
\alpha_1^2=\frac{b(c-d)}{c(b-d)},
\]
\[
g=\frac{2}{\sqrt{(a-c)(b-d)}},
\]
\[
\varphi=\sin^{-1}\sqrt{\frac{(b-d)(c-y)}{(c-d)(b-y)}},
\]
\[
u_1=F(\varphi,k),
\]
\[
k^2=\frac{(a-b)(c-d)}{(a-c)(b-d)},
\]
with $k^2<\alpha^2$.
For our specific choices in (\ref{abcdy}), these reduce to
\[
\alpha^2=\frac{2B}{A+B-1},
\]
\[
\alpha_1^2=\frac{2(A-1)}{A+B-1},
\]
\[
g=\frac{2B}{\sqrt{(A+B-1)(A-B+1)}},
\]
\[
k^2=\frac{4B}{(A+B-1)(A-B+1)},
\]
\[
\varphi=\frac{\pi}{2},
\]
and
\[
u_1=K(k).
\]
Specific cases include
\[
\int_y^c\frac{dx}{\sqrt{(a-x)(b-x)(c-x)(x-d)}}=gK(k)
\]
(Byrd \& Friedman (1954) \cite{ByrdFriedman}, (340.00)) and
\[
\int_y^c\frac{xdx}{\sqrt{(a-x)(b-x)(c-x)(x-d)}}=\frac{cg}{\alpha^2}
\left[
\alpha_1^2K(k)+
(\alpha^2-\alpha_1^2)\Pi(\alpha,k)
\right]
\]
(Byrd \& Friedman (1954) \cite{ByrdFriedman}, (340.01)).

In general we have
\[
\int_y^c\frac{x^pdx}{\sqrt{(a-x)(b-x)(c-x)(x-d)}}
=\frac{c^pg\alpha_1^{2p}p!}{\alpha^{2p}}
\sum_{j=0}^p\frac{(\alpha^2-\alpha_1^2)^j}{\alpha_1^{2j}j!(p-j)!}V_j
\]
(Byrd \& Friedman (1954) \cite{ByrdFriedman}, (340.04)), where
\[
\fl V_0=K(k),
\]
\[
\fl V_1=\Pi(\alpha,k),
\]
\[
\fl V_2=\frac{1}{2(\alpha^2-1)(k^2-\alpha^2)}
\left[ 
(k^2-\alpha^2)K(k)
+\alpha^2E(k)
+(2\alpha^2k^2+2\alpha^2-\alpha^4-3k^2)\Pi(\alpha,k)
\right],
\]
and larger values of $V_j$ can be computed using the following 
recurrence
relation
\begin{eqnarray}
V_{m+3}=\frac{1}{2(m+2)(1-\alpha^2)(k^2-\alpha^2)}\nonumber\\[0.2cm]
\hspace{1.8cm}\times\bigl[
(2m+1)k^2V_m
+2(m+1)(\alpha^2k^2+\alpha^2-3k^2)V_{m+1}\nonumber\\[0.2cm]
\hspace{3.60cm}+(2m+3)(\alpha^4-2\alpha^2k^2-2\alpha^2+3k^2)V_{m+2}
\bigr]\nonumber
\end{eqnarray}
(see Byrd \& Friedman (1954) \cite{ByrdFriedman}, (336.00--03)).
For instance,
\[
\eqalign{
\int_y^c\frac{x^2dx}{\sqrt{(a-x)(b-x)(c-x)(x-d)}}
\cr
\hspace{1cm}=\frac{c^2g}{\alpha^4}
\left[
\alpha_1^4K(k)+
2\alpha_1^2(\alpha^2-\alpha_1^2)\Pi(\alpha,k)+(\alpha^2-\alpha_1^2)^2V_2
\right]}
\]
(see Byrd \& Friedman (1954) \cite{ByrdFriedman}, (340.02)).

\medskip
In general, the Fourier coefficients for $\h^3$ will be given in terms of 
complete elliptic integrals of the first three kinds.
Let's directly compute the $m=0$ component,
in which (\ref{Amxtm}) reduces to
\[
{\displaystyle {\sf H}_0^{1/2}(r,r^\prime,\theta,\theta^\prime)=
-1+\frac{1}{\pi}}
{\displaystyle \int_{-1}^{1}\frac{\left(A/B-x\right)dx}
{\sqrt{
(1-x)(1+x)
\left(x-\frac{A+1}{B}\right)
\left(x-\frac{A-1}{B}\right)
}}}.
\]
Therefore using the above formulae, we have
\begin{eqnarray}
\fl\h_3|_{m=0}={\sf H}_0^{1/2}(r,r^\prime,\theta,\theta^\prime)\nonumber\\[0.2cm]
\hspace{-1.26cm}=-1+\frac{2K(k)}{\pi\sqrt{(A-B+1)(A+B-1)}}
+\frac{2(A-B-1)\Pi(\alpha,k)}{\pi\sqrt{(A-B+1)(A+B-1)}}\nonumber\\[0.2cm]
\hspace{-1.26cm}=-1+
\frac{2}{\pi}\left\{K(k)+\left[\cosh r\cosh r^\prime-\sinh r\sinh r^\prime\cos(\theta-\theta^\prime)-1\right]\Pi(\alpha,k)\right\}
\nonumber\\[0.0cm]
\hspace{2.6cm}\times\left[\cosh r\cosh r^\prime-\sinh r\sinh r^\prime\cos(\theta-\theta^\prime)+1\right]^{-1/2}\nonumber\\[0.2cm]
\hspace{2.6cm}\times\left[\cosh r\cosh r^\prime-\sinh r\sinh r^\prime\cos(\theta+\theta^\prime)-1\right]^{-1/2}
.\nonumber
\end{eqnarray}
Note that the Fourier coefficients 
\[
\h^3|_{m=0}\to\g^3|_{m=0},
\]
in the limit as $r,r^\prime\rightarrow 0^+$,
where $\g^3|_{m=0}$ is given in (\ref{gm0euc}).
This is expected since $\Hi_R^3$ is a manifold.

\subsection{Fourier expansion for a fundamental solution of the Laplacian on $\Hi_R^d$}
\label{FourierexpansionforaGreensfunctioninHid}

For the $d$-dimensional Riemannian manifold $\Hi_R^d$, with $d\ge 2$, one can expand
a fundamental solution of the Laplace-Beltrami operator in an azimuthal Fourier
series.  One may Fourier expand, in terms of the azimuthal coordinate, a fundamental
solution of the Laplace-Beltrami operator in any rotationally-invariant coordinate 
systems which admits solutions via separation of variables.  
In Euclidean space, there exist non-subgroup-type rotationally invariant coordinate 
systems which are separable for Laplace's equation.  All separable coordinate
systems for Laplace's equation in $d$-dimensional Euclidean space $\R^d$ are known.
In fact, this is also true for separable coordinate systems on $\Hi_R^d$ (see Kalnins (1986) 
\cite{Kalnins}).  There has been considerable work done in two and three dimensions, 
however there still remains a lot of work to be done for a detailed analysis of 
fundamental solutions.

We define an unnormalized fundamental solution of Laplace's equation 
on the unit hyperboloid
$\h^d:(\Hi^d\times\Hi^d)\setminus\{(\bfx,\bfx):\bfx\in\Hi^d\}\to\R$ such that
\begin{equation}
\h^d(\wbfx,\wbfxp):=\mcI_d(\rho(\wbfx,\wbfxp))=\frac{2\pi^{d/2}R^{d-2}}{\Gamma(d/2)}
\mch_R^d(\bfx,\bfxp).
\label{backhd}
\end{equation}
In our current azimuthal Fourier analysis, we therefore will focus on the relatively 
easier case of separable subgroup-type coordinate systems on $\Hi_R^d$, and specifically
for geodesic polar coordinates.  In these coordinates the Riemannian 
metric is given by (\ref{stanhypmetric}) and we further restrict our attention by 
adopting standard geodesic polar coordinates 
(\ref{standardhyp}).  

In these coordinates would would like to expand a
fundamental solution of Laplace's equation on the hyperboloid in an azimuthal 
Fourier series, namely
\begin{equation}
\fl\h^d(\wbfx,\wbfxp)=\sum_{m=0}^\infty \cos(m(\phi-\phi^\prime))
{\sf H}_m^{d/2-1}(r,r^\prime,\theta_1,\ldots,\theta_{d-2},\theta_1^\prime,\ldots,\theta_{d-2}^\prime)
\label{fulldFourierexpansion}
\end{equation}
where ${\sf H}_m^{d/2-1}:[0,\infty)^2\times[0,\pi]^{2d-4}\to\R$ is defined such that
%\[
%\mch_{R,m}^d =
%\mch_{R,m}^d
%(r,r^\prime,\theta_1,\ldots,\theta_{d-2},\theta_1^\prime,\ldots,\theta_{d-2}^\prime),
%\]
%for all $d\in\{2,3,\ldots\},$ $\bfx,\bfxp\in\Hi_R^d$, and $m\in\N_0$.  By standard Fourier 
%theory we know that
%\begin{equation}
%\mch_{R,m}^d=
%\frac{\epsilon_m}{\pi}\int_0^\pi \mch_R^d(\bfx,\bfxp) \cos(m(\phi-\phi^\prime)) d(\phi-\phi^\prime)
%\label{standardfouriertheoryhyperboloid}
%\end{equation}
\begin{equation}
\fl{\sf H}_m^{d/2-1}
(r,r^\prime,\theta_1,\ldots,\theta_{d-2},\theta_1^\prime,\ldots,\theta_{d-2}^\prime)
:=\frac{\epsilon_m}{\pi}\int_0^\pi \h^d(\wbfx,\wbfxp) \cos(m(\phi-\phi^\prime)) d(\phi-\phi^\prime)
\label{standardfouriertheoryhyperboloid}
\end{equation}
(see for instance Cohl \& Tohline (1999) \cite{CT}).
According to Theorem \ref{thmh1d} and (\ref{backhd}), we may write 
$\h^d(\wbfx,\wbfxp)$ in terms 
of associated Legendre functions of the second kind as follows
\begin{equation}
\fl\h^d(\wbfx,\wbfxp)=\frac{e^{-i\pi(d/2-1)}}{2^{d/2-1}\Gamma(d/2)\,(\sinh d(\wbfx,\wbfxp))^{d/2-1}}
Q_{d/2-1}^{d/2-1}\left(\cosh d(\wbfx,\wbfxp)\right).
\label{Qrepfundsolhyperboloid}
\end{equation}
By (\ref{dhyperboloid}) we know that in any geodesic polar
coordinate system 
\begin{equation}
\cosh d(\wbfx,\wbfxp)=\cosh r \cosh r^\prime-\sinh r\sinh r^\prime\cos\gamma,
\label{coshdxxp}
\end{equation}
and therefore through (\ref{standardfouriertheoryhyperboloid}),
(\ref{Qrepfundsolhyperboloid}), and
(\ref{coshdxxp}),
in standard geodesic polar coordinates,
the azimuthal Fourier coefficient can be given by
\begin{equation}
\hspace{-1.5cm} \eqalign{{\sf H}_m^{d/2-1}(r,r^\prime,\theta_1,\ldots,\theta_{d-2},\theta_1^\prime,\ldots,\theta_{d-2}^\prime)\cr
\hspace{2.5cm}=\frac{\epsilon_m e^{-i\pi(d/2-1)}}{2^{d/2-1}\pi\Gamma(d/2)}
\int_0^\pi
\frac{Q_{d/2-1}^{d/2-1}\left(A-B\cos\psi\right)\cos(m\psi)}
{\left[(A-B\cos\psi)^2-1\right]^{(d-2)/4}}d\psi,}
\label{compactexpressionforfouriercoeff}
\end{equation}
where $\psi:=\phi-\phi^\prime,$ $A,B:[0,\infty)^2\times[0,\pi]^{2d-4}\to\R$ 
are defined through
(\ref{prodform}) and (\ref{coshdxxp}) 
as
\[
\fl\displaystyle A(r,r^\prime,\theta_1,\ldots,\theta_{d-2},\theta_1^\prime,\ldots,\theta_{d-2}^\prime):=
\cosh r\cosh r^\prime
\sum_{i=1}^{d-2}\cos\theta_i{\cos\theta_i}^\prime
\prod_{j=1}^{i-1}\sin\theta_j{\sin\theta_j}^\prime,
\]
and
\[
\fl\displaystyle
B(r,r^\prime,\theta_1,\ldots,\theta_{d-2},\theta_1^\prime,\ldots,\theta_{d-2}^\prime):=
\sinh r\sinh r^\prime
\prod_{i=1}^{d-2}\sin\theta_i{\sin\theta_i}^\prime.
\]

Even though 
(\ref{compactexpressionforfouriercoeff}) is a compact expression for the Fourier
coefficient of a fundamental solution of Laplace's equation on $\Hi_R^d$ for $d\in\{2,3,4,\ldots\},$ 
it may be informative to use any of the representations of a fundamental solution
of the Laplacian on $\Hi_R^d$ from Theorem \ref{thmh1d} to express the Fourier coefficients.
For instance if one uses the finite-summation expression in the odd-dimensions, on can write
the Fourier coefficients as a linear combination of integrals of the form
\[
\int_{-1}^1\frac
{\left[(a+b)/2-x\right]^{2k-1}x^pdx}
{(a-x)^{k-1}(b-x)^{k-1}\sqrt{(a-x)(b-x)(c-x)(x-d)}},
\]
where $x=\cos\psi$, $k\in\{1,\ldots,(d-1)/2\}$, $p\in\{0,\ldots,m\},$
and we have used the nomenclature of section \ref{FourierexpansionforaGreensfunctioninHi3}.
This integral is a rational function of $x$ multiplied by an inverse square-root of 
a quartic in $x$.  Because of this and due to the limits of integration, we see 
that by definition, these are all given in terms of complete elliptic integrals.
The special functions which represent the azimuthal Fourier coefficients on $\Hi_R^d$
are unlike the odd-half-integer degree, integer-order, associated Legendre functions 
of the second kind which 
appear in Euclidean space $\R^d$ for $d$ odd (see 
Cohl (2010) \cite{CohlthesisII};
Cohl \& Dominici (2010) \cite{CohlDominici}), in that 
they include complete elliptic integrals of the third kind 
(in addition to complete elliptic integrals of the first and second kind)
(cf.~section \ref{FourierexpansionforaGreensfunctioninHi3})
in their basis functions.
For $d\ge 2$, through
(4.1) in Cohl \& Dominici (2010) \cite{CohlDominici} and that $\Hi_R^d$ is a 
manifold (and therefore must locally represent Euclidean space), the
functions ${\sf H}_m^{d/2-1}$ are generalizations of associated Legendre functions of 
the second kind with odd-half-integer degree and order given by either 
an odd-half-integer or an integer.

\section{Gegenbauer expansion in geodesic polar coordinates}
\label{Gegenbauerexpansioninhyperbolichypersphericalcoordinates}

\noindent In this section we derive an eigenfunction expansion for a fundamental solution 
of Laplace's equation on the hyperboloid in geodesic polar 
coordinates for $d\in\{3,4,\ldots\}.$  Since the spherical harmonics for $d=2$ are just trigonometric functions 
with argument given in terms of the azimuthal angle,
this case has already been covered in section \ref{FourierexpansionforaGreensfunctioninHi2}.\\[-0.2cm]

\noindent In geodesic polar coordinates, Laplace's equation is given by (cf.~(\ref{genhyplap}))
\begin{equation}
\Delta f=\frac{1}{R^2}\left[\frac{\partial^2f}{\partial r^2}
+(d-1)\coth r\frac{\partial f}{\partial r}
+\frac{1}{\sinh^2r} \Delta_{\Si^{d-1}}\right]f=0,
\label{eq5}
\end{equation}
where $f:\Hi_R^d\to\R$ and $\Delta_{\Si^{d-1}}$ is the corresponding 
Laplace-Beltrami operator on the $(d-1)$-dimensional unit sphere $\Si^{d-1}$. 
Eigenfunctions $Y_l^K:\Si^{d-1}\to\C$ of the Laplace-Beltrami operator $\Delta_{\Si^{d-1}}$,
where $l\in\N_0$ and $K$ is a set of quantum numbers 
which label representations for $l$ in separable subgroup type coordinate systems on $\Si^{d-1}$
(i.e.\ angular momentum type quantum numbers, see Izmest'ev {\it et al.}~(2001) \cite{IPSWb}),
are given by solutions to the eigenvalue problem
(\ref{eq4}).

In standard geodesic polar coordinates (\ref{standardhyp}), 
$K=(k_1,\ldots,k_{d-3},|k_{d-2}|)\in\N_0^{d-2}$ with 
$k_0=l\ge k_1 \ge \ldots \ge k_{d-3} \ge |k_{d-2}| \ge 0$, and in particular
$k_{d-2}\in\{-k_{d-3},\ldots,k_{d-3}\}$.
A positive fundamental solution $\mch_R^d:(\Hi_R^d\times\Hi_R^d)\setminus\{(\bfx,\bfx):\bfx\in\Hi_R^d\}\to\R$
on the $R$-radius hyperboloid satisfies (\ref{eq3}).
The completeness relation for hyperspherical harmonics in standard hyperspherical coordinates is given by
\[
\sum_{l=0}^\infty\sum_K
Y_l^K(\theta_1,\ldots,\theta_{d-1})
\overline{Y_l^K(\theta_1^\prime,\ldots,\theta_{d-1}^\prime)} =
\frac{\delta(\theta_1-\theta_1^\prime)\ldots\delta(\theta_{d-1}-\theta_{d-1}^\prime)}
{\sin^{d-2}\theta_{d-1}^\prime\ldots\sin\theta_2^\prime},
\]
where $K=(k_1,\ldots,k_{d-2})$ and $l=k_0\in\N_0$.
Therefore through (\ref{diracdeltasubg}), we can write
\begin{equation}
\delta_g(\bfx,\bfxp)=\frac{\delta(r-r^\prime)}{R^d\sinh^{d-1}r^\prime}
\sum_{l=0}^\infty\sum_K
Y_l^K(\theta_1,\ldots,\theta_{d-1})
\overline{Y_l^K(\theta_1^\prime,\ldots,\theta_{d-1}^\prime)}.
\label{eq1}
\end{equation}
For fixed $r,r^\prime\in[0,\infty)$ and $\theta_1^\prime,\ldots,\theta_{d-1}^\prime\in[0,\pi]$,
since $\mch_R^d$ is harmonic on its domain, its restriction is in $C^2(\Si^{d-1})$,
and therefore has a unique expansion in hyperspherical harmonics, namely 
\begin{equation}
\mch_R^d(\bfx,\bfxp)=\sum_{l=0}^\infty\sum_K u_l^K(r,r^\prime,\theta_1^\prime,\ldots,\theta_{d-1}^\prime)
Y_l^K(\theta_1,\ldots,\theta_{d-1}),
\label{eq2}
\end{equation}
where $u_l^K:[0,\infty)^2\times[0,\pi]^{d-1}\to\C$.
If we substitute (\ref{eq1}) and (\ref{eq2}) into (\ref{eq3}) and use (\ref{eq5}) and (\ref{eq4}), we obtain
\begin{eqnarray}
\fl\sum_{l=0}^\infty\sum_K
Y_l^K(\theta_1,\ldots,\theta_{d-1})
\left[
\frac{d^2}{dr^2}
+(d-1)\coth r\frac{d}{dr}
-\frac{l(l+d-2)}{\sinh^2r}\right]
u_l^K(r,r^\prime,\theta_1^\prime,\dots,\theta_{d-1}^\prime) 
\nonumber\\[-0.0cm]
\hspace{0.2cm}
=\sum_{l=0}^\infty\sum_K
Y_l^K(\theta_1,\ldots,\theta_{d-1})
\overline{Y_l^K(\theta_1^\prime,\ldots,\theta_{d-1}^\prime)}\cdot
\frac{\delta(r-r^\prime)}{R^{d-2}\sinh^{d-1}r^\prime}.
\label{sumprodylkulkprimes}
\end{eqnarray}
This indicates that for $u_l:[0,\infty)^2\to\R$,
\begin{equation}
u_l^K(r,r^\prime,\theta_1^\prime,\ldots,\theta_{d-1}^\prime)=u_l(r,r^\prime)
\overline{Y_l^K(\theta_1^\prime,\ldots,\theta_{d-1}^\prime)},
\label{simplifyulk}
\end{equation}
and from (\ref{eq2}) the expression for a fundamental of the
Laplace-Beltrami operator in hyperspherical coordinates on the hyperboloid is given by
\begin{equation}
\mch_R^d(\bfx,\bfxp)=\sum_{l=0}^\infty u_l(r,r^\prime)
\sum_{K}
Y_l^K(\theta_1,\ldots,\theta_{d-1})
\overline{Y_l^K(\theta_1^\prime,\ldots,\theta_{d-1}^\prime)}.
\label{sumulprodylks}
\end{equation}
The above expression can be simplified using 
the addition theorem for hyperspherical 
harmonics (see for instance Wen \& Avery (1985) \cite{WenAvery}, section 10.2.1 in Fano \& Rau (1996), 
Chapter 9 in Andrews, Askey \& Roy (1999) \cite{AAR} and especially 
Chapter XI in Erd{\'e}lyi {\it et al.}~Vol.~II (1981) \cite{ErdelyiHTFII}),
which is given by 
\begin{equation}
\sum_{K}
Y_l^K
%(\theta_1,\ldots,\theta_{d-1})
(\wbfx)
\overline{Y_l^K
%(\theta_1^\prime,\ldots,\theta_{d-1}^\prime)
(\wbfxp)
}
=\frac{\Gamma(d/2)}{2\pi^{d/2}(d-2)}
(2l+d-2)
C_l^{d/2-1}(\cos\gamma),
\label{additionthmhypspheharm}
\end{equation}
where $\gamma$ is the angle between two arbitrary 
vectors 
$\wbfx,\wbfxp\in\Si^{d-1}$
given in terms of (\ref{cosgamma}).
The Gegenbauer polynomials $C_l^\mu:[-1,1]\to\R$, $l\in\N_0$, $\mbox{Re}\,\mu>-1/2$,
can be defined in terms of the Gauss hypergeometric function as 
\[
C_l^\mu(x):=\frac{(2\mu)_l}{l!}\,{}_2F_1\left(-l,l+2\mu;\mu+\frac12;\frac{1-x}{2}\right).
\]
The above expression (\ref{sumulprodylks})
can be simplified using 
(\ref{additionthmhypspheharm}), therefore
\begin{equation}
\mch_R^d(\bfx,\bfxp)=
\frac{\Gamma(d/2)}{2\pi^{d/2}(d-2)}
\sum_{l=0}^\infty u_l(r,r^\prime)
(2l+d-2) C_l^{d/2-1}(\cos\gamma).
\label{almostfinalhdexpression}
\end{equation}

Now we compute the exact expression for $u_l(r,r^\prime)$.
By separating the angular
dependence in (\ref{sumprodylkulkprimes}) and using (\ref{simplifyulk}),
we obtain the differential equation
\begin{equation}
\frac{d^2 u_l}{dr^2}
+(d-1)\coth r\frac{du_l}{dr}
-\frac{l(l+d-2)u_l}{\sinh^2r} = -\frac{\delta(r-r^\prime)}{R^{d-2}\sinh^{d-1}r^\prime}.
\label{eq7}
\end{equation}
Away from $r=r^\prime$, solutions to the differential equation (\ref{eq7}) 
must be given by solutions to the homogeneous equation, which are are given 
in section \ref{Laplaceequationandharmonicsonthehyperboloid}.  Therefore, 
the solution to (\ref{eq7}) is given by
\begin{equation}
\fl u_l(r,r^\prime)=\frac{A}{\left(\sinh r\sinh r^\prime\right)^{d/2-1}}
P_{d/2-1}^{-(d/2-1+l)}(\cosh r_<)
Q_{d/2-1}^{d/2-1+l}(\cosh r_>),
\label{eq8}
\end{equation}
such that $u_l(r,r^\prime)$ is continuous at $r=r^\prime$, 
where $A\in\R.$

In order to determine the constant $A$, we first make the substitution
\begin{equation}
v_l(r,r^\prime)=(\sinh r\sinh r^\prime)^{(d-1)/2}u_l(r,r^\prime).
\label{substitutionuv}
\end{equation}
This converts (\ref{eq7}) into the following differential equation
\[
\fl\frac{\partial^2v_l(r,r^\prime)}{\partial r^2}
-\frac14
\left[
\frac{(d-1+2l)(d-3+2l)}{\sinh^2r}+(d-1)^2
\right]v_l(r,r^\prime)=-\frac{\delta(r-r^\prime)}{R^{d-2}},
\]
which we then integrate over $r$ from $r^\prime-\epsilon$
to $r^\prime + \epsilon$, and take the limit as $\epsilon\to 0^+$.  
We are left with a discontinuity condition for 
the derivative of $v_l(r,r^\prime)$ with respect to $r$ evaluated at $r=r^\prime$, namely
\begin{equation}
\lim_{r\to 0^+}\left.\frac{dv_l(r,r^\prime)}{dr}\right\vert
_{r^\prime-\epsilon}
^{r^\prime+\epsilon}
=\frac{-1}{R^{d-2}}.
\label{eq9}
\end{equation}
After inserting (\ref{eq8}) with (\ref{substitutionuv})
into (\ref{eq9}), substituting $z=\cosh r^\prime$,
evaluating at $r=r^\prime$, and making use of the Wronskian relation (e.g.~p.~165 in 
Magnus, Oberhettinger \& Soni (1966) \cite{MOS})
\[
W\left\{P_\nu^{-\mu}(z),Q_\nu^\mu(z)\right\}=-\frac{e^{i\pi\mu}}{z^2-1},
\]
which is equivalent to
\[
W\left\{P_\nu^{-\mu}(\cosh r^\prime),Q_\nu^\mu(\cosh r^\prime)\right\}=-\frac{e^{i\pi\mu}}{\sinh^2r^\prime},
\]
we obtain
\[
A=\frac{e^{-i\pi(d/2-1+l)}}{R^{d-2}},
\]
and hence
\[
\fl u_l(r,r^\prime)=\frac{e^{-i\pi(d/2-1+l)}}{R^{d-2}(\sinh r\sinh r^\prime)^{d/2-1}}
P_{d/2-1}^{-(d/2-1+l)}(\cosh r_<)
Q_{d/2-1}^{d/2-1+l}(\cosh r_>),
\]
and therefore through (\ref{almostfinalhdexpression}), we have
\begin{eqnarray}
\fl\mch_R^d(\bfx,\bfxp)=
\frac{\Gamma(d/2)}{2\pi^{d/2}R^{d-2}(d-2)}
\frac{e^{-i\pi(d/2-1)}}{(\sinh r\sinh r^\prime)^{d/2-1}}
\nonumber\\[0.0cm]
\hspace{-1.7cm}\times\sum_{l=0}^\infty 
(-1)^l
(2l+d-2) 
P_{d/2-1}^{-(d/2-1+l)}(\cosh r_<)Q_{d/2-1}^{d/2-1+l}(\cosh r_>)
C_l^{d/2-1}(\cos\gamma).
\label{gegenexpansion}
\end{eqnarray}

As an alternative check of our derivation, we can do the asymptotics for the product of
associated Legendre functions
$P_{d/2-1}^{-(d/2-1+l)}(\cosh r_<)Q_{d/2-1}^{d/2-1+l}(\cosh r_>)$ 
in (\ref{gegenexpansion})
as $r,r^\prime\to 0^+$.
The appropriate asymptotic
expressions for $P$ and $Q$ respectively can be found on p.~171 and p.~173 in Olver (1997) \cite{Olver}.
For the associated Legendre function of the first kind there is
\[
P_\nu^{-\mu}(z)\sim\frac{\left[(z-1)/2\right]^{\mu/2}}{\Gamma(\mu+1)},
\]
as $z\to 1$, $\mu\ne -1,-2,\ldots$,  and for the associated Legendre function of the
second kind there is
\[
Q_\nu^{\mu}(z)\sim\frac
{e^{i\pi\mu}\Gamma(\mu)}
{2\left[(z-1)/2\right]^{\mu/2}},
\]
as $z\to 1^+$, $\mbox{Re}\ \,\mu>0$, and $\nu+\mu\ne -1,-2,-3,\ldots$.  To second order the
hyperbolic cosine is given by $\cosh r\simeq 1+r^2/2$. Therefore to lowest order
we can insert $\cosh r_<\simeq 1+r_<^2/2$ and $\cosh r_>\simeq 1+r_>^2/2$ into the above expressions 
yielding
\[
P_{d/2-1}^{-(d/2-1+l)}(\cosh r_<)\sim\frac{(r_</2)^{d/2-1+l}}{\Gamma(d/2+l)},
\]
and
\[
Q_{d/2-1}^{d/2-1+l}(\cosh r_>)\sim\frac{e^{i\pi(d/2-1+l)}\Gamma(d/2-1+l)}{2(r_>/2)^{d/2-1+l}},
\]
as $r,r^\prime\to 0^+$.
Therefore the asymptotics for the product of associated Legendre functions in
(\ref{gegenexpansion}) is given by
\begin{equation}
P_{d/2-1}^{-(d/2-1+l)}(\cosh r_<)
Q_{d/2-1}^{d/2-1+l}(\cosh r_>)\sim
\frac{e^{i\pi(d/2-1+l)}}{2l+d-2}
\left(\frac{r_<}{r_>}\right)^{l+d/2-1}
\label{prodasymptotics}
\end{equation}
(the factor $2l+d-2$ is a term which one encounters regularly with hyperspherical harmonics).
Gegenbauer polynomials obey the following generating function
\begin{equation}
\frac{1}
{\left(1+z^2-2zx\right)^\mu}=\sum_{l=0}^\infty C_l^\mu(x) z^l,
\label{genfuncgegen}
\end{equation}
\noindent where $x\in[-1,1]$ and $|z|<1$
(see for instance, p.~222 in Magnus, Oberhettinger \& Soni (1966) \cite{MOS}).
The generating function for Gegenbauer polynomials (\ref{genfuncgegen})
can be used to expand a fundamental solution of Laplace's equation 
in Euclidean space $\R^d$ (for $d\ge 3$, 
cf.~Theorem \ref{thmg1n}) in hyperspherical coordinates, namely
\begin{equation}
\frac{1}{\|\bfx-\bfxp\|^{d-2}}=\sum_{l=0}^\infty \frac{r_<^l}{r_>^{l+d-2}}
C_l^{d/2-1}(\cos\gamma),
\label{eucfundsolnExpansion}
\end{equation}
where $\gamma$ was defined in (\ref{additionthmhypspheharm}).
Using (\ref{eucfundsolnExpansion}) and Theorem \ref{thmg1n}, 
since $\mch_R^d\to\mcg^d$, $\sinh r,\sinh r^\prime\to r,r^\prime$ 
and (\ref{prodasymptotics}) is satisfied to lowest order as $r,r^\prime\to 0^+$,
we see that
(\ref{gegenexpansion})
obeys the correct asymptotics 
and
our fundamental solution expansion locally reduces 
to the appropriate expansion for Euclidean space, as it should since $\Hi_R^d$ is a manifold.

Note that (\ref{gegenexpansion})
can be further expanded over the remaining $(d-2)$-quantum numbers in $K$ 
in terms of a simply separable product of normalized harmonics 
$Y_l^K(\wbfx)\overline{Y_l^K(\wbfxp)}$,
where $\wbfx,\wbfxp\in\Si^{d-1}$,
using the addition theorem for hyperspherical harmonics (\ref{additionthmhypspheharm})
(see Cohl (2010) \cite{CohlthesisII} for several examples).

It is intriguing to observe how one might obtain the Fourier expansion 
for $d=2$ (\ref{expansiondequals2}) from the the expansion 
(\ref{gegenexpansion}), which is strictly valid for $d \ge 3$.
If one makes the substitution $\mu=d/2-1$ in
(\ref{gegenexpansion}) then we obtain the following conjecture
(which matches up to the generating
function for Gegenbauer polynomials in the Euclidean limit $r,r^\prime\to 0^+$)
\begin{eqnarray}
\fl\frac{1}{\sinh^\mu\rho}Q_\mu^\mu(\cosh\rho)&=&
\frac{2^\mu\Gamma(\mu+1)}{(\sinh r\sinh r^\prime)^\mu}\nonumber\\[0.2cm]
&&\hspace{0.2cm}\times\sum_{n=0}^\infty (-1)^n\frac{n+\mu}{\mu}P_\mu^{-(\mu+n)}(\cosh r_<)Q_\mu^{\mu+n}(\cosh r_>)
C_n^\mu(\cos\gamma),
\label{conj}
\end{eqnarray}
for all $\mu\in\C$ such that $\mbox{Re}\,\mu>-1/2$.
If we take the limit as $\mu\to 0$ in (\ref{conj}) and use
\begin{equation}
\lim_{\mu\to 0}\frac{{n}+\mu}{\mu}C_{n}^\mu(x)=\epsilon_{n} T_{n}(x)
\label{limitcheby}
\end{equation}
(see for instance (6.4.13) in Andrews, Askey \& Roy (1999) \cite{AAR}),
where $T_{n}:[-1,1]\to\R$ is the Chebyshev polynomial of the first kind
defined as $T_l(x):=\cos(l\cos^{-1}x),$ then we obtain the following formula
\[
\fl\frac{1}{2}\log\frac{\cosh\rho+1}{\cosh\rho-1}=
\sum_{n=0}^\infty\epsilon_n(-1)^nP_0^{-n}(\cosh r_<)
Q_0^n(\cosh r_>)\cos(n(\phi-\phi^\prime)),
\]
where $\cosh\rho=\cosh r\cosh r^\prime-\sinh r\sinh r^\prime\cos(\phi-\phi^\prime)$.
By taking advantage of the following formulae
\begin{equation}
P_0^{-n}(z)=\frac{1}{n!}\left[\frac{z-1}{z+1}\right]^{n/2},
\label{Pzerominusn}
\end{equation}
for $n\ge 0,$ 
\begin{equation}
Q_0(z)=\frac12\log\frac{z+1}{z-1}
\label{Qzero}
\end{equation}
(see (8.4.2) in Abramowitz \& Stegun (1972) \cite{Abra}), and
\begin{equation}
Q_0^n(z)=\frac12 (-1)^n(n-1)!\left\{
\left[
\frac{z+1}{z-1}
\right]^{n/2}
-
\left[
\frac{z-1}{z+1}
\right]^{n/2}
\right\},
\label{Qzeron}
\end{equation}
for $n\ge 1$ then 
(\ref{expansiondequals2}) is reproduced.
The representation (\ref{Pzerominusn}) follows easily from the Gauss hypergeometric
representation of the associated Legendre function of the first kind (see (8.1.2) in 
Abramowitz \& Stegun (1972) \cite{Abra})
\begin{equation}
P_\nu^\mu(z)=\frac{1}{\Gamma(1-\mu)}\left[\frac{z+1}{z-1}\right]^{\mu/2}
{}_2F_1\left(-\nu,\nu+1;1-\mu;\frac{1-z}{2}\right).
\label{associatedlegendreP}
\end{equation}
One way to derive the representation of the associated Legendre function of second kind 
(\ref{Qzeron}) is to use Whipple formula for associated Legendre functions (cf.~(8.2.7)
in Abramowitz \& Stegun (1972) \cite{Abra})
\[
Q_\nu^\mu(z)=\sqrt{\pi}{2}\Gamma(\nu+\mu+1)(z^2-1)^{-1/4}e^{i\pi\mu}P_{-\mu-1/2}^{-\nu-1/2}
\left(\frac{z}{\sqrt{z^2-1}}\right),
\]
and (8.6.9) in Abramowitz \& Stegun (1972) \cite{Abra}, namely
\[
\fl P_\nu^{-1/2}(z)=\sqrt{\frac{2}{\pi}}\frac{(z^2-1)^{-1/4}}{2\nu+1}
\left\{
\left[
z+\sqrt{z^2-1}
\right]^{\nu+1/2}
-\left[
z+\sqrt{z^2-1}
\right]^{-\nu-1/2}
\right\},
\]
for $\nu\neq -1/2$.

\subsection{Addition theorem for the azimuthal Fourier coefficient on $\Hi_R^3$}
\label{AdditiontheoremorazimuthalFouriercoefficientsonHiR3}

One can compute addition theorems for the azimuthal Fourier coefficients of
a fundamental solution for Laplace's equation on $\Hi_R^d$ for $d\ge 3$ by 
relating directly obtained Fourier coefficients to the expansion over 
hyperspherical harmonics for the same fundamental solution.  By using the
expansion of $\mch_R^d(\bfx,\bfxp)$ in terms of Gegenbauer polynomials 
(\ref{gegenexpansion})
in combination with the addition theorem for hyperspherical harmonics
(\ref{additionthmhypspheharm}) expressed in, for instance, one of Vilenkin's 
polyspherical coordinates (see section IX.5.2 in Vilenkin (1968) \cite{Vilen};
Izmest'ev {\it et al.}~(1999,2001) \cite{IPSWa,IPSWb}),
one can obtain through series rearrangement a multi-summation expression for 
the azimuthal Fourier coefficients. Vilenkin's polyspherical coordinates are
simply subgroup-type coordinate systems which parametrize points on $\Si^{d-1}$
(for a detailed discussion of these coordinate systems see 
chapter 4 in Cohl (2010) \cite{CohlthesisII}).  In this section we will
give an explicit example of just such an addition theorem on $\Hi_R^3$.

The azimuthal Fourier coefficients on $\Hi_R^3$ expressed in standard hyperspherical
coordinates (\ref{standardhyp}) are given by the functions  
${\sf H}_m:[0,\infty)^2\times[0,\pi]^2\to\R$ which is defined by 
(\ref{azimuthalFourierintegralh3}).  By expressing 
(\ref{gegenexpansion}) in $d=3$ we obtain
\begin{eqnarray}
&&\hspace{-2cm}\mch_R^3(\bfx,\bfxp)=\frac{-i}{4\pi R\sqrt{\sinh r\sinh r^\prime}}\nonumber\\[0.2cm]
&&\hspace{0.4cm}\times\sum_{l=0}^\infty (-1)^l(2l+1)P_{1/2}^{-(1/2+l)}(\cosh r_<)
Q_{1/2}^{1/2+l}(\cosh r_>) P_l(\cos\gamma),
\label{HR3expansiongeg}
\end{eqnarray}
where $P_l:[-1,1]\to\R$ is the Legendre polynomial defined by $P_l(x)=C_l^{1/2}(x)$,
or through (\ref{associatedlegendreP}) with $\mu=0$ and $\nu\in\N_0$.  By using 
the addition theorem for hyperspherical harmonics (\ref{additionthmhypspheharm}) 
with $d=3$ using $(\cos\theta,\sin\theta\cos\phi,\sin\theta\sin\phi)$ to parametrize
points on $\Si^2,$ we have since the normalized spherical harmonics are
\[
Y_{l,m}(\theta,\phi)=(-1)^m\sqrt{\frac{2l+1}{4\pi}\frac{(l-m)!}{(l+m)!}}P_l^m(\cos\theta)e^{im\phi},
\]
the addition theorem for spherical harmonics, namely
\begin{equation}
P_l(\cos\gamma)=\sum_{m=-l}^l\frac{(l-m)!}{(l+m)!}P_l^m(\cos\theta)P_l^m(\cos\theta^\prime)
e^{im(\phi-\phi^\prime)},
\label{addnthmhypsph}
\end{equation}
%where $\cos\gamma=\cos\theta\cos\theta^\prime+\sin\theta\sin\theta^\prime\cos(\phi-\phi^\prime)$.
where $\cos\gamma$ is given by (\ref{cosgammah3}).
By combining (\ref{HR3expansiongeg}) and (\ref{addnthmhypsph}), reversing the order of the
two summation symbols, and comparing the result with (\ref{azimuthalfourierexpansionh3})
we obtain the following single summation addition theorem for the azimuthal Fourier coefficients
of a fundamental solution of Laplace's equation on $\Hi_R^3$, namely since $\h^3=4\pi R\mch_R^3$,
\begin{eqnarray*}
&&\hspace{-2.0cm}
{\sf H}_m^{1/2}(r,r^\prime,\theta,\theta^\prime)
=\frac{-i\epsilon_m}{\sqrt{\sinh r\sinh r^\prime}}
\sum_{l=|m|}^\infty (-1)^l(2l+1)\frac{(l-m)!}{(l+m)!}
\nonumber\\[0.2cm]
&&\hspace{2.0cm}\times
P_l^m(\cos\theta)
P_l^m(\cos\theta^\prime)
P_{1/2}^{-(1/2+l)}(\cosh r_<)
Q_{1/2}^{1/2+l}(\cosh r_>).
\end{eqnarray*}
This addition theorem reduces to the corresponding result 
((2.4) in Cohl {\it et al.} (2001) \cite{CRTB})
in the Euclidean $\R^3$ limit as $r,r^\prime\to 0^+$.

\section{Discussion}
\label{Discussion}

Re-arrangement of the multi-summation expressions in
section \ref{Gegenbauerexpansioninhyperbolichypersphericalcoordinates}
is possible through modification of the order in which the 
countably infinite space of quantum numbers is summed over in a standard 
hyperspherical coordinate system, namely
\begin{eqnarray}
\sum_{l=0}^\infty
\sum_{K}=
\sum_{l=0}^\infty
\sum_{k_1=0}^l
\sum_{k_2=0}^{k_1}
\cdots
\sum_{k_{d-4}=0}^{k_{d-5}}
\sum_{k_{d-3}=0}^{k_{d-4}}
\sum_{k_{d-2}=-k_{d-3}}^{k_{d-3}}\nonumber\\[0.0cm]
\hspace{1.40cm}=
\sum_{k_{d-2}=-\infty}^{\infty}
\sum_{k_{d-3}=|k_{d-2}|}^\infty
\sum_{k_{d-4}=k_{d-2}}^\infty
\cdots
\sum_{k_2=k_3}^\infty
\sum_{k_1=k_2}^\infty
\sum_{k_0=k_1}^\infty.\nonumber
\end{eqnarray}
Similar multi-summation re-arrangements have been accomplished previously for azimuthal
Fourier coefficients of fundamental
solutions for the Laplacian in Euclidean space (see for instance
Cohl {\it et al.} (2000) \cite{CTRS}; Cohl {\it et al.} (2001) \cite{CRTB}).
Comparison of the azimuthal Fourier expansions 
in section \ref{FourierexpansionsforaGreensfunctioninthehyperboloidmodel} 
(and in particular 
(\ref{compactexpressionforfouriercoeff}))
with re-arranged Gegenbauer expansions
in section \ref{Gegenbauerexpansioninhyperbolichypersphericalcoordinates}
(and in particular
(\ref{gegenexpansion}))
will yield new addition theorems for the special functions representing the
azimuthal Fourier coefficients of a fundamental solution of the Laplacian on
the hyperboloid.
These implied addition theorems will provide new special function identities
for the azimuthal Fourier coefficients, which are
hyperbolic generalizations of particular associated Legendre functions of the second kind.
In odd-dimensions, these special functions reduce to toroidal harmonics.

%A similar analysis of fundamental solutions for Laplace's equation, such as that given
%in this paper, is possible in all rank one symmetric spaces (see for instance
%Helgason (1984) \cite{Helgason84}).  The 

\section*{Acknowledgements}
Much thanks to A.~Rod Gover, Tom ter Elst, Shaun Cooper, and 
Willard Miller, Jr.~for valuable discussions.  I would like to express my 
gratitude to Carlos Criado Camb\'{o}n in the Facultad de Ciencias at 
Universidad de M\'{a}laga for his assistance in describing the global 
geodesic distance function in the hyperboloid model.  
We would also like to acknowledge two anonymous referees whose comments
helped improve this paper.  I acknowledge 
funding for time to write this paper from the Dean of the Faculty of 
Science at the University of Auckland in the form of a three month 
stipend to enhance University of Auckland 2012 PBRF Performance.
Part of this work was conducted while H.~S.~Cohl was a National Research Council
Research Postdoctoral Associate in the Information Technology Laboratory at the 
National Institute of Standards and Technology, Gaithersburg, Maryland, U.S.A.

\section*{References}
%\bibliography{/home/hcohl/tex/refbib.bib}

\end{document}